\documentclass[a4,12pt]{article}
\usepackage{graphics}
\usepackage{epsfig}
\usepackage{epsf}

\begin{document}
\title{Heterogeneity and Increasing Returns May Drive Socio-Economic Transitions}

\author{G{\'e}rard~Weisbuch$^{*}$, Vincent Buskens$^{**}$, Luat Vuong$^{***}$
\\[1cm]
$^{*}$Laboratoire de Physique Statistique\footnote{Laboratoire associ{\'e} au
CNRS (UMR 8550), {\`a} l'ENS et aux Universit{\'e}s Paris 6 et Paris 7.} de l'Ecole Normale Sup{\'e}rieure, \\
    24 rue Lhomond, F-75231 Paris Cedex 5, France. \\
 {\small {\em email}: weisbuch@lps.ens.fr}\\
$^{**}$Department of Sociology, Utrecht University, \\ Heidelberglaan 2,
3584 CS, Utrecht, The Netherlands. \\ {\small{\em email}: v.buskens@uu.nl}\\
$^{***}$School of Applied and Engineering Physics,
Cornell University, \\
Ithaca, NY 14853 USA \\
}

\maketitle

\abstract{\noindent There are clear benefits associated with a particular
consumer choice for many current markets. For example, as we consider here,
some products might carry environmental or `green' benefits. Some consumers
might value these benefits while others do not. However, as evidenced by myriad
failed attempts of environmental products to maintain even a niche market, such
benefits do not necessarily outweigh the extra purchasing cost. The question we
pose is, how can such an initially economically-disadvantaged green product
evolve to hold the greater share of the market? We present a simple
mathematical model for the dynamics of product competition in a heterogeneous
consumer population. Our model preassigns a hierarchy to the products, which
designates the consumer choice when prices are comparable, while prices are
dynamically rescaled to reflect increasing returns to scale. Our approach
allows us to model many scenarios of technology substitution and provides a
method for generalizing market forces. With this model, we begin to forecast
irreversible trends associated with consumer dynamics as well as 
policies that could be made to influence transitions.}

\section{Introduction}
\label{intro}


The adoption of `green' technology is hard to predict because it implies many
intertwined social and economic factors plus many retro-action loops. Because
even a partial description of these factors and their interaction seems so
intricate, authors are attracted by the multi-agent modelling approach (see,
e.g., \cite{jasss} and the related special issue of JASSS). But this approach
suffers many limitations, especially when it comes to obtain a full description
of the dynamics in the parameter space or to get some insight in what happens
in the model and how it can be compared to the real modelled system. This is
especially important because one might observe sharp changes in predictions in
very small areas of the parameter space (see also \cite{weisbuch2000}).

We here propose a simple soluble model that takes into account some of the
intricacies of real life problems such as the heterogeneity in consumer
responses, social influence, and increasing returns to scale of production
prices but still condenses the set of parameters that can lead to unwieldy
complexity in simulation (see also \cite{kemp} for a related model, but without
increasing returns to scale). The increasing returns to scale could also be
interpreted as increasing social pressure to buy a specific product if most
people around you do this, an argument already made by
B. Arthur\cite{art} . To maintain our analysis on the competition between
goods with different prices and different levels of environmental soundness, we
subsume that the influences of different cultures, social influences,
government policy, advertising can all be summarized into a single curve, which
we name the `willingness to pay' (WTP) function. WTP illuminates the consumer
population distribution as a function of price and describes how people vary in
their extent to which they want to pay for environmental benefits. We assume
this normalized WTP distribution has a fixed shape, while the price associated
with each market product varies with its market share.

Our model has three ancillary assumptions. Firstly, given that there are two
products with prices below what a consumer is willing to pay, this consumer
chooses the `greener' technology. The first assumption implies that we can
assume without further loss of generality also that environmentally superior
technologies are more expensive given a similar market share. Given the first
assumption, products that are more expensive and less `green' cannot
survive in this market. Secondly, prices decrease with market shares. The
combination of buyers heterogeneity in WTP and increasing returns results in a
rich variety of dynamical regimes with ultimately different prices and market
shares. Thirdly, we assume that while each product varies in maximum price,
they each experience a similar extent of linear decreasing returns to scale.
This last assumption could very easily be relaxed in future extensions of the
model, but is not crucial for the implications we derive in this paper.

In our investigation we focus our attention on a market competition between
three cars: a standard (0), hybrid (1), `green' (2). Figure~1 shows a
bell-shaped WTP distribution where the prices associated with each car, $p_0$,
$p_1$, and $p_2$ determine the corresponding market share areas, $u_0$, $u_1$,
and $u_2$, shaded in red, blue, and green, respectively. Consumers with WTP
larger than $p_2$ choose car~2. Of the remaining market share, consumers with
WTP larger than $p_1$ choose car~1. Finally, from the still remaining market
share, consumers with WTP larger than $p_0$ choose car 0. According to the WTP
distribution, some agents might decide to buy no car so that $u_0 + u_1 + u_2
\leq 1$.

This paper is organized as follows. Section~\ref{assumptions} describes our
computational model with equations. Section~\ref{results} describes our
results. We show that environmental technologies will take over 
when a significant  fraction of agents have already a WTP
to pay for the green car whatever its market share and  
when the increasing returns coefficient is large (because of strong social
effects or large production price reduction with production level). We also
demonstrate how different conditions can lead to car~0, 1, or 2 dominating the
market depending on the rescaling of price due to demand.
Subsection~\ref{regimes} shows that our simple model not only provides reliable
results but also analytic tools to evaluate multiple asymptotic solutions. For
some parameter values, the dynamics have several attractors, which implies
hysteresis effects. We extrapolate that the timing of the subsidies and grants
at the immediate onset or emergence of a new technology may be crucial as trends
are sometimes irreversible. Section~\ref{conclusion} summarizes our conclusions
and shows future direction for this model.

\section{A Simple Set of Assumptions}
\label{assumptions}

We start the description of our model with explaining in more detail our
assumptions. We use the example of more or less green cars below, but clearly
any product for which there is some non-monetary benefit related to the product
can be used instead. So where we write for convenience `car' below, one could
read `product,' and when we write environmental benefits or `greenness,' one
could read more non-monetary benefits (ammenities).

\begin{itemize}

\item Consumers care about two aspects of a car, i.e., the price and
`greenness' of the car;

\item Environmentally superior technologies are more expensive (given a similar
market share);

\item Cars with a larger market share are cheaper;

\item People vary in their extent to which they want to pay for environmental
benefits (heterogeneity of WTP) \cite{gonaph, gonaphjv,naphgo}

\item People choose the alternative they prefer.

\end{itemize}

The second assumption is more for convenience than that it is really necessary.
Environmentally inferior technologies that are more expensive would not be
chosen by anybody, because consumers only care about these two aspects.

The third assumption has also two possible interpretations. A larger market
share has advantages of scale for the producers, which implies that cars can be
produced for a lower price when the market share is larger. From the buyers
perspective, social influence effects affect the utility consumers get from a
product: a more popular car is more attractive for most individuals than a car
that nobody bought, which changes the specific willingness to pay for that type
of car. But this is equivalent to saying that the price of that type of car
decreases. Since only the difference between WTP and prices is relevant to
buyers' decisions, both effects have the same results \cite{art}. In theory one
should make a distinction between the production costs which are influenced by
market shares say during the actual year, and the social influence terms which
takes into account how many cars of each type have been bought in the past (the
integrated yearly market share). In practice, because of the irreversibility of
the capital investment in the automobile industry, the actual production cost
also integrate investments and the two effects are driven by the integrated
market share.

There are three technological options: standard (0), intermediate (1), `green'
(2).

\begin{itemize}

\item Each option $i$ has its own maximum cost $P_{0i}$, which equals the cost
at a zero market share;

\item All options have the same linear returns to scale coefficient $k$ (i.e.,
$p_i = P_{0i} - k u_{i}$, where $p_i$ and $u_{i}$ are the actual price and market
share of product $i$, respectively;)

\item There is one distribution of WTP for environmental benefits (e.g.,
uniform or bell-shaped)

\end{itemize}

\begin{figure}[htbp]
\centerline{\epsfxsize=120mm\epsfbox{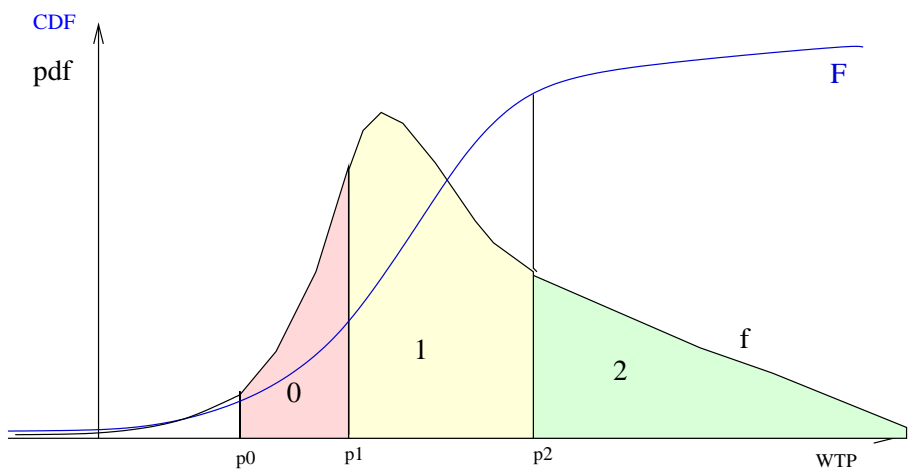}}

\caption{Distribution of willingness to pay (WTP) and market shares. Consumers
with WTP larger than $p_i$ may choose car $i$ with market share $u_{i}$ equal
to the colored area.}
\end{figure}

Figure 1 shows an arbitrary distribution of WTP. Consumers with WTP larger than
$p_i$ may choose car $i$ with market share $u_{i}$ equal to the colored area.
In the following subsection, we explain how the stable market shares are
computed.

\subsection{Equations}

Given that we assume the same returns to scale, the price of a product follows
a linear return to scale function.

\begin{equation}
  p_i = P_{0i} - k u_{i}
\end{equation}

Because of the influence of market share on actual prices, the order among
actual prices $p_i$ may differ from maximum prices at zero market share. A
first operation is to order products $i$. New indices $j$ are used for prices
ordered by increasing actual prices, according to market share distribution.
Some products might lose their rank: in such a case, we consider that every
time a product $j$ has larger price $p_j$ than a product
with a better environmental quality, it disappears from the market: nobody is
interested to buy a more expensive product with a lower environmental quality.
With final indices $j$, equilibrium market shares obey:

\begin{equation}
  u_j = F(p_{j+1}) - F(p_{j})
\end{equation}
where $F(p)$ is the cumulative WTP distribution. This simple set of equations
is soluble, either directly for simple expressions of $F(p)$, as we do in
section 3.2 on results or through transcendental equations.

While these equations provide the (possible multiple) solutions for the stable
proportions of each product in the market, one can also simulate the dynamics
to these solution by assuming that at each point in time a certain proportion
$\lambda$ of consumers is choosing a new car according to the prices and
preferences at that point in time. The related dynamics of market shares is
given by:
\begin{equation}
  u_j(t+1) = (1-\lambda) u_{j}(t)+ \lambda ( F(p_{j+1},t) - F(p_{j},t))
\end{equation}

\begin{figure}[htbp]
\centerline{\epsfxsize=120mm\epsfbox{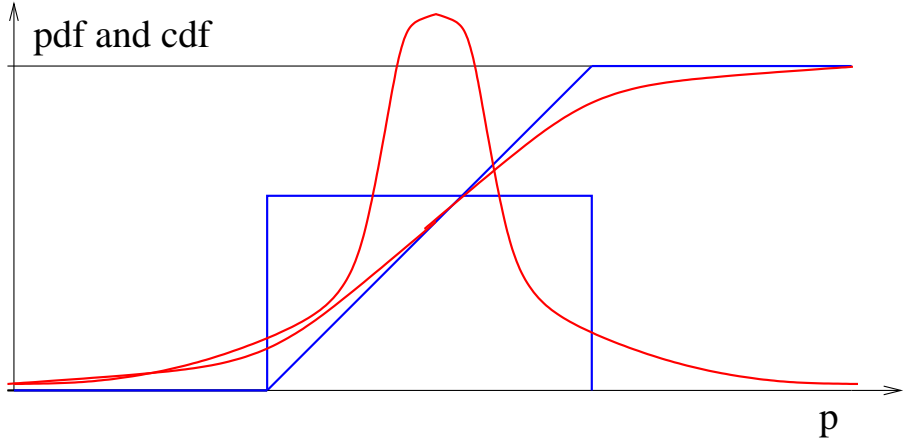}} \caption{Distributions of
willingness to pay (WTP). A uniform partial distribution function (pdf) and the
corresponding cumulative distribution function (cdf) are drawn in black. The
corresponding logit distributions are drawn in red (bell-shaped for the pdf,
S-shaped for the cdf).}
\end{figure}

The simulations in this paper were done for two simple WTP distributions: the
uniform distribution and a logit distribution. The corresponding equation for
the cumulative distribution $F(p)$ is piecewise linear for the uniform
distribution:
\begin{equation}
  F(p)= \frac{p-p_m}{p_M-p_m}
\end{equation}
between the minimum $p_m$  and the maximum $p_M$ WTP price. Below $p_m$,
$F(p)=0$, above $p_M$, $F(p)=1$.

For the bell-shaped cumulative distribution we use a logit expression:
\begin{equation}
   F(p)= \frac{1}{1+ exp(-\beta p)}
\end{equation}
where $\beta$ is inversely proportional to the width of the distribution. If we
define the width as the inverse slope at the point of inflexion ($p=0$), we
obtain:
 \begin{equation}
    w = \frac{4}{\beta}.
 \end{equation}

The corresponding graphs appear in figure 2.

\section{Results}
\label{results}

\subsection{Time evolution of market shares}
\label{evolution}

The evolution of market shares for the three products is easily simulated for a
fixed set of parameters. The two plots of figure 3 only differ by the value of
the maximum price of the green product $P_{02}$. In both cases, as in most
simulations, only the standard product is initially present
($u_0(t=0)=1,u_1(0)=0, u_2(0)=0$). This corresponds with the assumption that we
want to predict at which maximum prices it is possible to enter the market for
greener products.

Asymptotic market shares are reached in a few characteristic times
$(\lambda)^{-1}$, where $(\lambda)^{-1}$ corresponds to a minimum
characteristic time of evolution towards equilibrium.

The connection between the evolution of market shares and prices is evident
from figure 4, obtained for the same parameter values as the right plot of
figure 3. The initial increase of $u_1$ and $u_2$ decreases $u_0$, $p_1$, and
$p_2$, and increases $p_0$. Since $p_1$ decreases slower than $p_2$, $u_1$
saturates after an initial increase and finally decreases. It is also clear
that due to the lower $P_{02}$ in the right plot, the green car becomes the
dominant product in the market and completely drives out again the intermediate
car, while the standard car maintains a minority share in the market

Because of the increase of $p_0$, some consumers do not find any car to buy
(unless $k$ is larger than the width of the distribution, see further the
section on hysteresis). The asymptotic sum of market shares is then less than
one. Such a situation is often encountered.

\begin{figure}[htbp]
\centerline{\epsfxsize=100mm\epsfbox{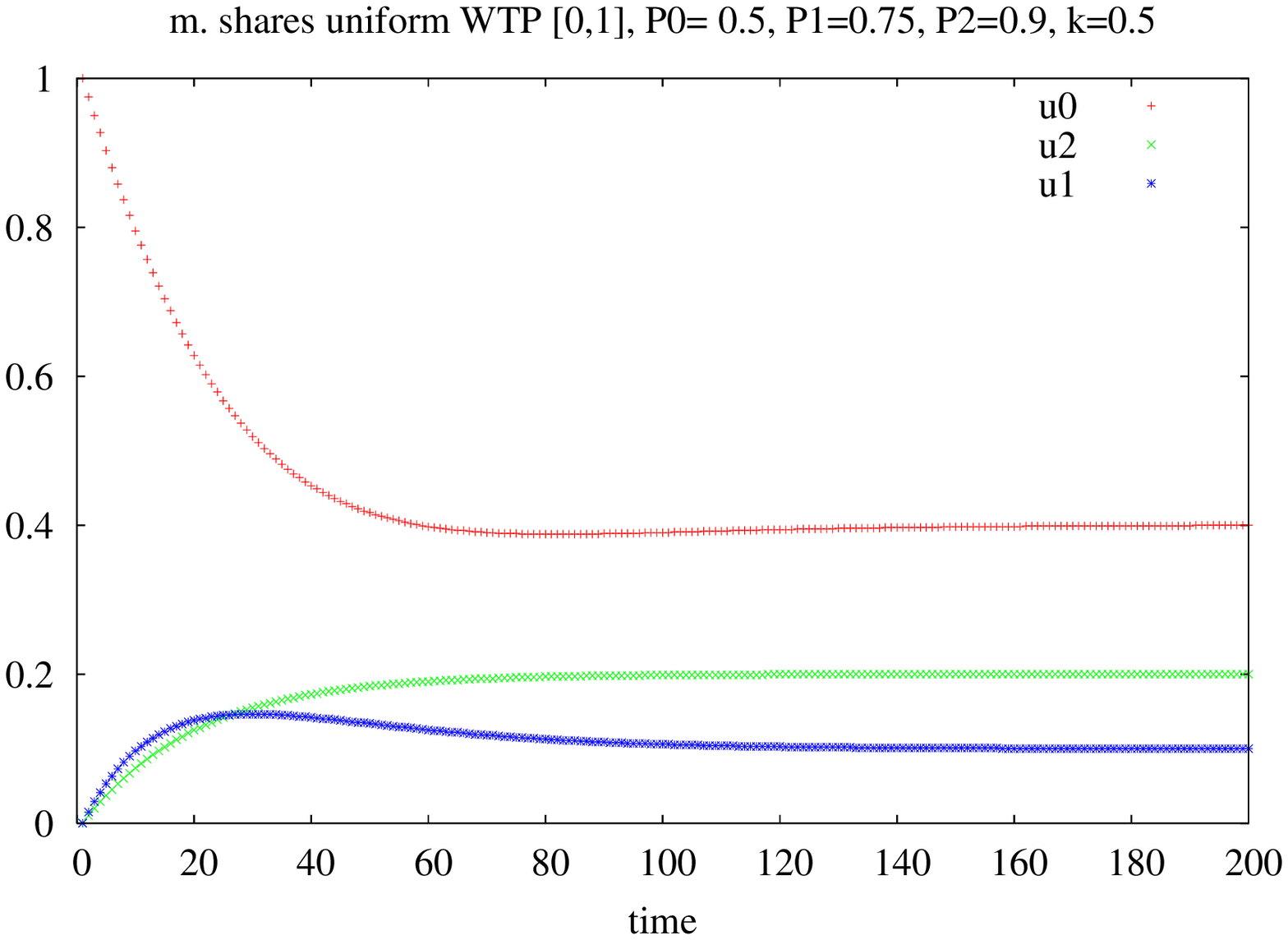} \epsfxsize=100mm\epsfbox{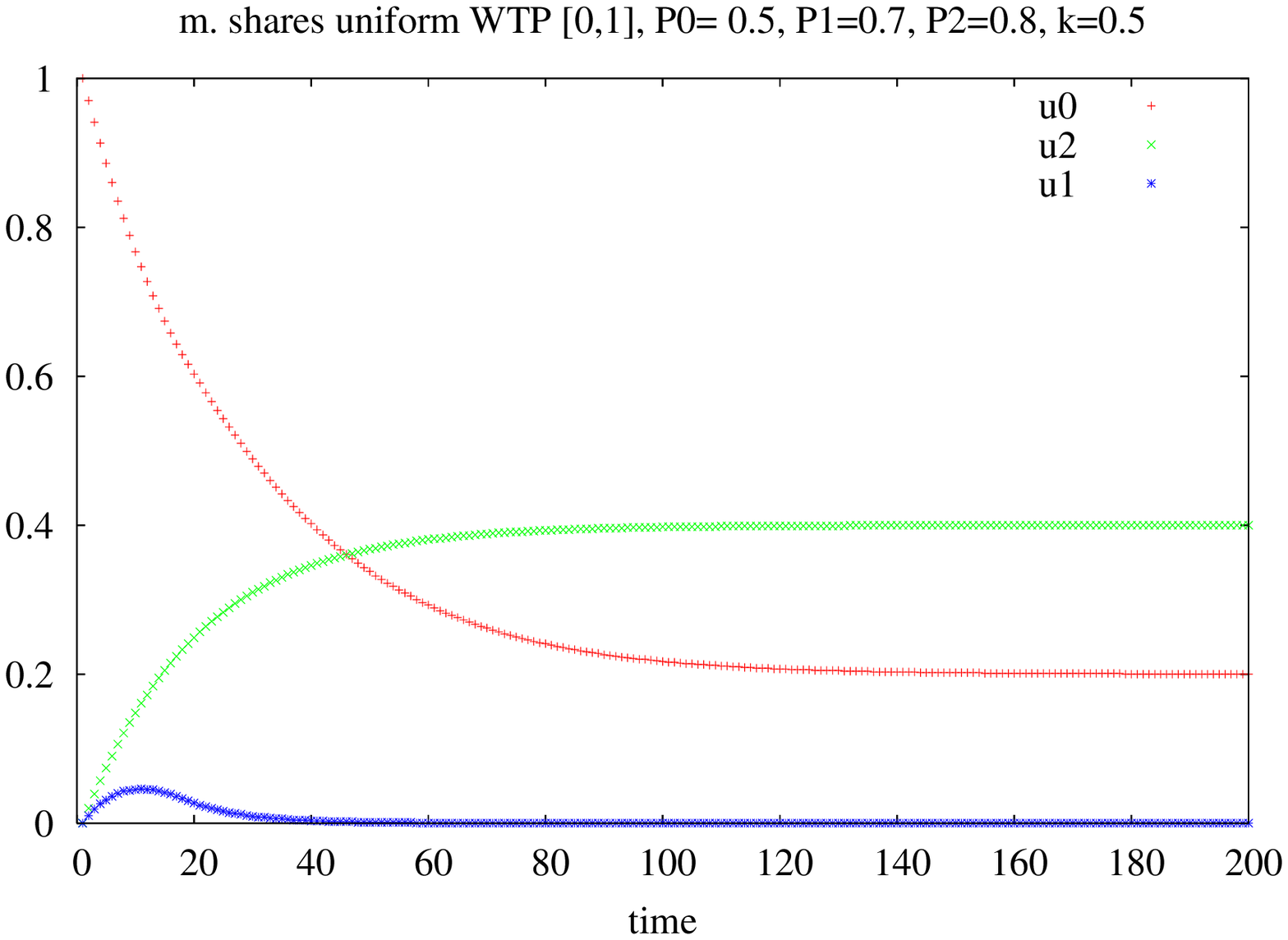}}
\caption{Time evolutions of market shares: red standard, blue
  intermediate, green `green' products. $(\lambda)^{-1}=10$. }
\end{figure}

\begin{figure}[htbp]
\centerline{\epsfxsize=120mm\epsfbox{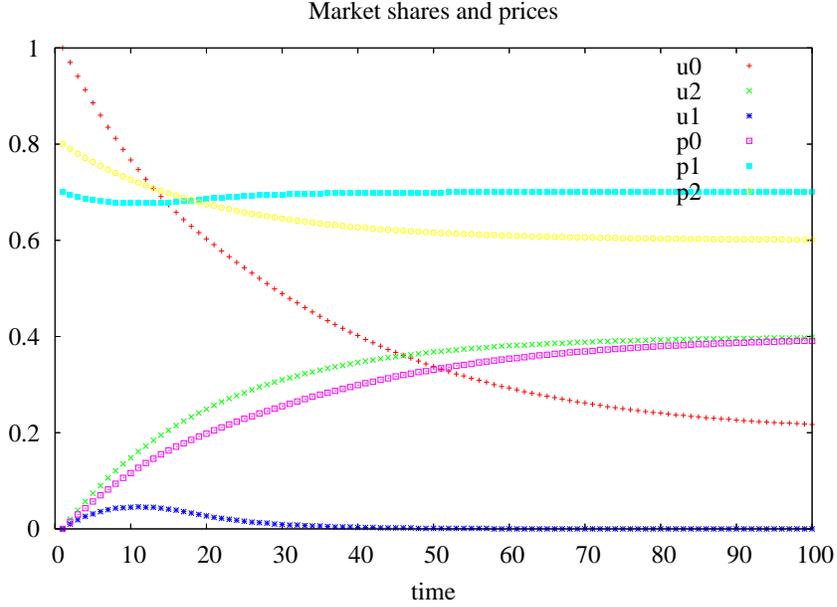}}
\caption{Time evolutions of market shares
and prices. }
\end{figure}

\subsection{Dynamical regimes}
\label{regimes}

To obtain a more complete overview about how the parameters affect the outcomes
of the dynamics, we will now provide a more complete overview of possible end
points of the dynamics, which we label `dynamical regimes.'  The parameters are
a priori:

\begin{itemize}

\item Two parameters defining the center of the WTP distribution and its width
(we only study symmetric distributions).

\item The three maximum prices $P_{0i}$.

\item The slope $k$ of the increasing returns expression.
\end{itemize}

They in fact reduce to four independent parameters, because only the relative
position of prices with respect to the WTP distribution is important. We will
find that only the ratio of $k$ to the width of the WTP distributions plays a
role for simple distributions.

One should realize that $\lambda$ is a kinetic parameter that only influences
how fast attractors are reached, but not the attractors themselves.

The following figures were made for constant maximum prices of standard and
green product, and for fixed WTP. We sometimes used uniform, and sometimes the
logit distribution. The two varying parameters are then $k$ and $P_{01}$. This
turns out to provide already a quite complete overview of which dynamical
regimes occur and how they depend on the parameters.

\begin{figure}[htbp]
\centerline{\epsfxsize=120mm\epsfbox{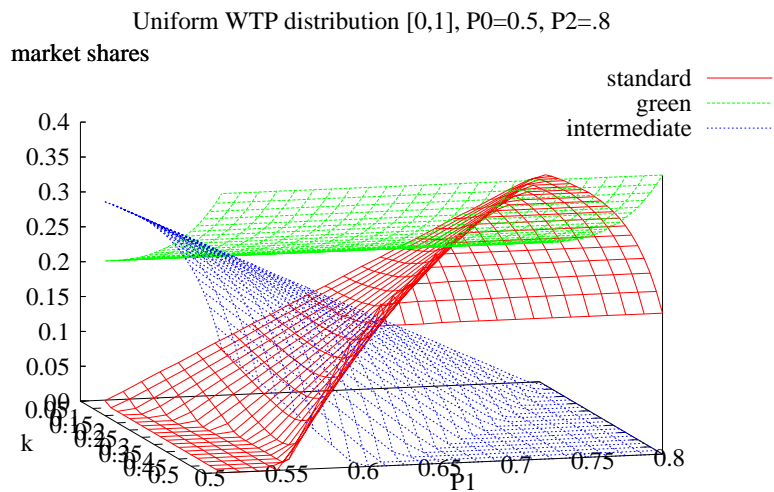}}

\caption{Asymptotic market shares as of function of $k$ and $P_{01}$. Uniform
WTP distribution [0,1], $P_{00}=0.5$, $P_{02}=0.8$. $k$ varies between 0 (no
return to scale) and 0.5. $P_{01}$ varies between $P_{00}$ and $P_{02}$. Green,
blue, and red corresponds to the market share of the green, intermediate, and
standard products, respectively. }
\end{figure}

\begin{figure}[htbp]
\centerline{\epsfxsize=120mm\epsfbox{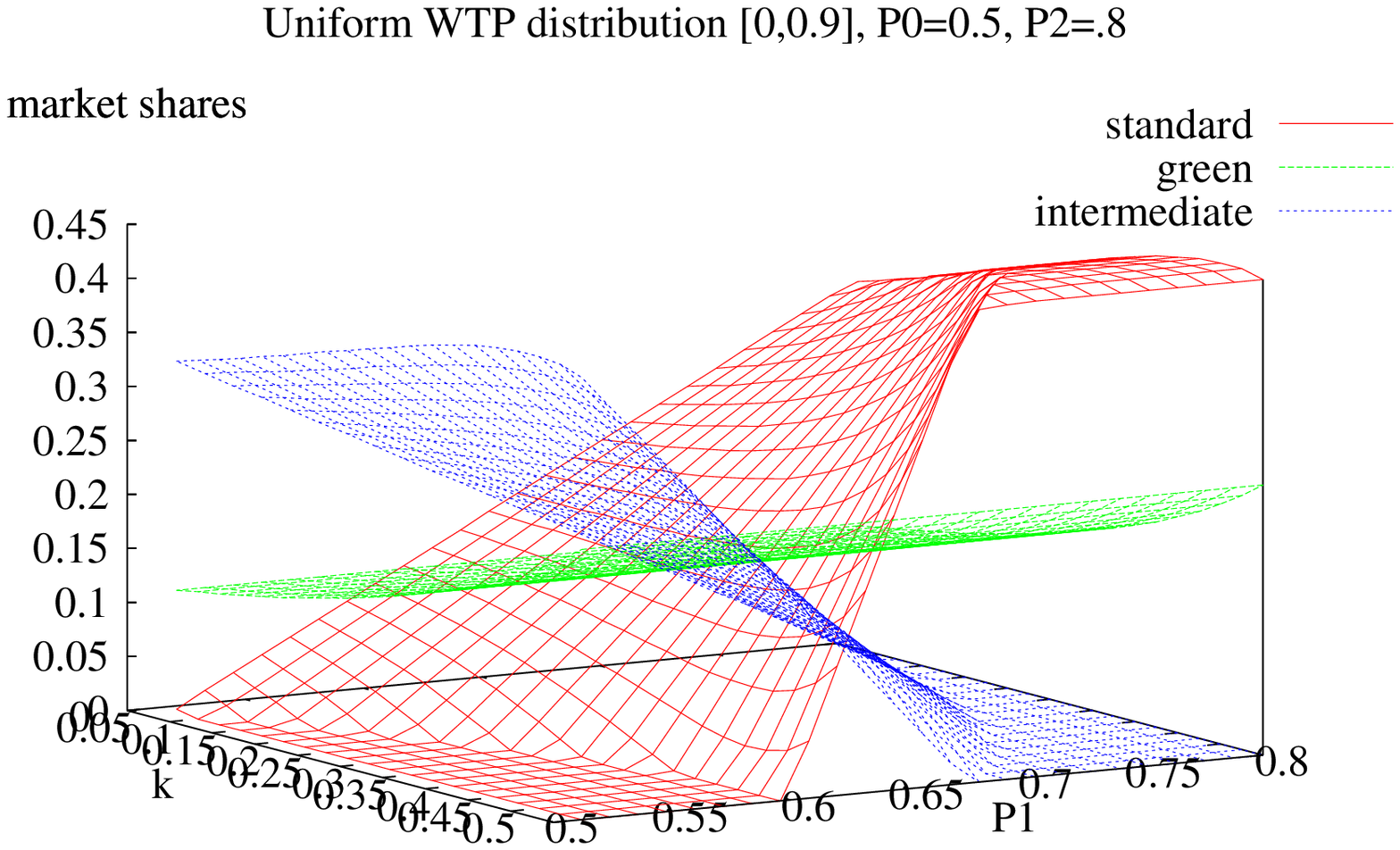}}
\caption{Asymptotic market shares as of function of $k$ and $P_{01}$.
 Uniform WTP distribution [0,0.9].
 All other items as in figure 5.}
\end{figure}

\begin{figure}[htbp]
\centerline{\epsfxsize=120mm\epsfbox{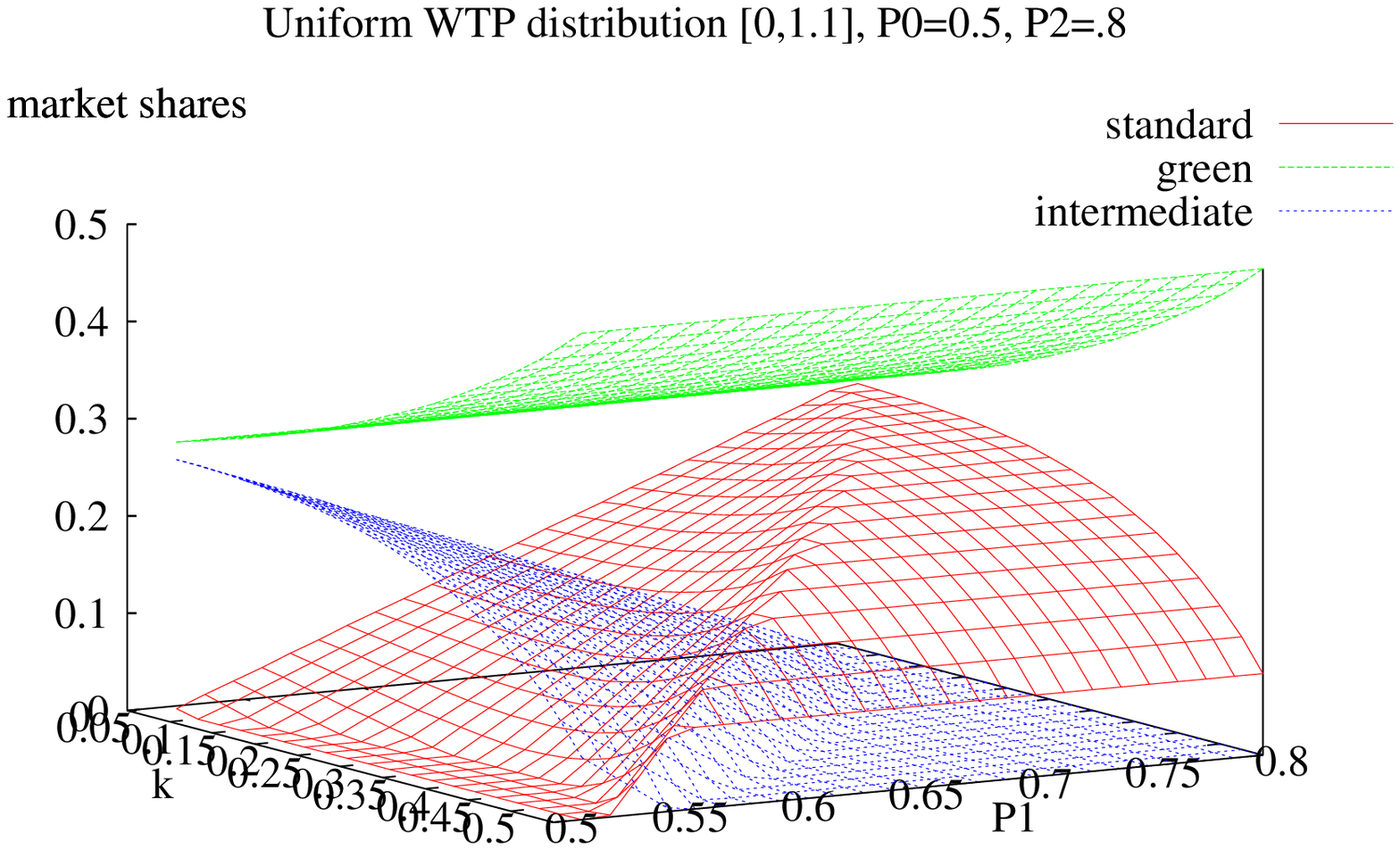}}
\caption{Asymptotic market shares as of function of $k$ and $P_{01}$.
 Uniform WTP distribution [0,1.1].
 All other items as in the figure 5.}
\end{figure}

For the above choice of parameters, whether the green product becomes dominant
depends essentially from how far $P_{02}$ is from the maximum WTP: in other
words what is the potential market share taking only into account the maximum
price of the green product. Of course, $u_2$ increases with $k$. In fact, $u_2$
is independent of $P_{01}$ and $P_{00}$. At equilibrium it always obeys:
\begin{equation}
  u_2 = 1 - F(P_{02} - ku_2)
\end{equation}
  since the `green' product is always chosen when the customer's WTP is larger
than $p_2$.

  For the uniform WTP distribution this gives:
\begin{equation}
  u_2 =  \frac{p_M-P_{02}}{p_M-p_m-k}
\end{equation}
  where $p_M$ and $p_m$ are respectively the upper and lower bound of the WTP
distribution. This expression is only valid for $k < w$ (check the hysteresis
section for the opposite case). In addition, $p_M$ should be larger than
$P_{02}$ otherwise $u_2 = 0$. It shows that the market share of the green car
only depends upon two reduced parameters, $\frac{p_M-P_{02}}{p_M-p_m}$ and
$\frac{k}{p_M-p_m}$.

The competition between the standard and intermediate car, on the other hand,
mostly depends upon $P_{01}$: the standard car is favored when $P_{01}$ is
close to $P_{02}$, and the intermediate car when $P_{01}$ is close to $P_{00}$.

For the uniform WTP distribution, equations (2) in $u_2$, $u_1$, and $u_0$ are
easily solved; the expressions for $u_2$, $u_1$, and $u_0$ depend of the
ranking of prices. In the case of $p_0 \le p_1 \le p_2$, they are written in
the simple case of a [0,1] WTP distribution:
  \begin{eqnarray}
     u_2 = \frac{1 - P_{02}}{1-k}\\
       u_1 = \frac{p_2 - P_{01}}{1-k}\\
         u_0 = \frac{p_1 - P_{00}}{1-k}
  \end{eqnarray}
  where actual prices $p_2$ and $p_1$ are obtained from the corresponding
values of $u_i$. One sees from these equations that markets shares are zero
(and the equations have to be re-ordered) whenever $p_{i+1} \le P_{0i}$. These
conditions re-written in terms of initial parameters are:
\begin{eqnarray}
  u_1=0 \quad \mbox{iff} \quad P_{01} \ge \frac{P_{02}-k}{1-k}\\
  u_0=0 \quad \mbox{iff} \quad \mbox{either } P_{00} \ge \frac{P_{01}}{1-k} -
  \frac{k(P_{02}-k)}{(1-k)^2} \mbox{ and } P_{01} < \frac{P_{02}-k}{1-k}\\
  \mbox{           or} \quad P_{00} \ge \frac{P_{02} - k}{1-k} \mbox{ and } P_{01} \ge \frac{P_{02}-k}{1-k}
\end{eqnarray}

For more general distributions, e.g., the logit distribution, a similar
iterative procedure yields market shares and prices, but the equations are
transcendental rather than explicitly soluble.

\subsection{Hysteresis}

Equations such as equation (2) are well-known in physics, e.g., in the Mean
Field theory of ferromagnetism \cite{kit} 
 and to some extent in economics \cite{fol,fish}, in
the case of increasing returns or social influence. They are known to produce
`phase' or `regime' transitions when the number of their solutions goes from
one to three as a function of a parameter which is $k/w$ in our case (for the
logit distribution $w=4/\beta$).

Figure 8 allows to easily understand the regime transition. It is a graphical
solution of equation (2) rewritten as:
\begin{equation}
 1-u = F(p-ku).
\end{equation}
The curves corresponding to the two sides of the equation are drawn in the plan
$(p-ku, 1-u)$. The red and black curves correspond to $F(p-ku)$ (respectively,
logit and uniform distributions). The blue lines correspond to the left hand
side $1 - u$ as a function of $p - ku$; it is a straight line, which abscissa
is $p$ when $u=0$ and $p-k$ when $u=1$. Two blue lines are drawn corresponding
to the two cases of a large $k$, $k_M$ and a small $k$, $k_m$.

\begin{figure}[htbp]
\centerline{\epsfxsize=120mm\epsfbox{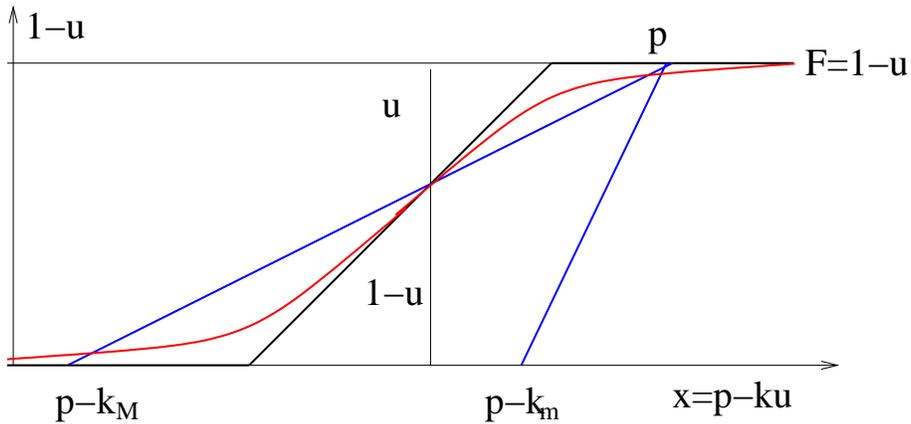}} \caption {Graphical
solution of equation (2) rewritten as $1-u = F(p-ku)$. Abscissa is $x= p-ku$,
and ordinate $1-u$. The $F(x)$ curve (cdf), represented in red for the logit
distribution of WTP and in black for a uniform distribution, intersects blue
lines $1-u$ in one or three points according to the value of $k$ for the
particular choice of $p$. $k_m \le 1 \le k_M$.}
\end{figure}

Three solutions may be obtained if the slope of the straight line $1-u$ is
lower than the largest slope of the $F$ function. This slope is $k/w$ for the
uniform distribution and $4k/\beta$ for the logit distribution. The
corresponding conditions for $k$ versus the WTP distribution parameters are
thus written:
\begin{equation}
   k  \ge  w
\end{equation}
\begin{equation}
  k  \ge  \frac{4}{\beta}
\end{equation}

When $k$ is above the threshold, one or three fixed points are then obtained
depending upon the value of the maximum price. In the case of three fixed
points, the central one is unstable but the two extreme are attractors; which
attractor is actually reached depends upon initial conditions: 
large $u$ values attractor is, e.g., obtained when initially $u(0)=1$ 
(and small $u$ attractor when
$u(0)=0$). The central fixed point separates the two attractor regions.

We considered until now $P_{02}$ as a fixed parameter. Let us now discuss the
effect of variations of $P_{02}$, which could be due to technological progress
or to strategic moves of the producer of car 2, or to government subsidies etc.
The two next figures illustrate the hysteresis cycle when $P_{02}$ is varied,
and the transition between the two regimes when both $k$ and $P_{02}$ are
varied. They were drawn by superimposing asymptotic simulation results obtained
when initial conditions were first $u_2(0)=1$ (red curves) and then when
$u_2(0)=0$ (green curves). In the first case, we thus start without standard or
intermediate cars. In the second case, we start with only standard cars. For
parameters such that only one attractor exists, only the green curves are
visible. But when two are possible, the red curves are visible and correspond
to $u_2(0) = 1$.

\begin{figure}[htbp]
\centerline{\epsfxsize=120mm\epsfbox{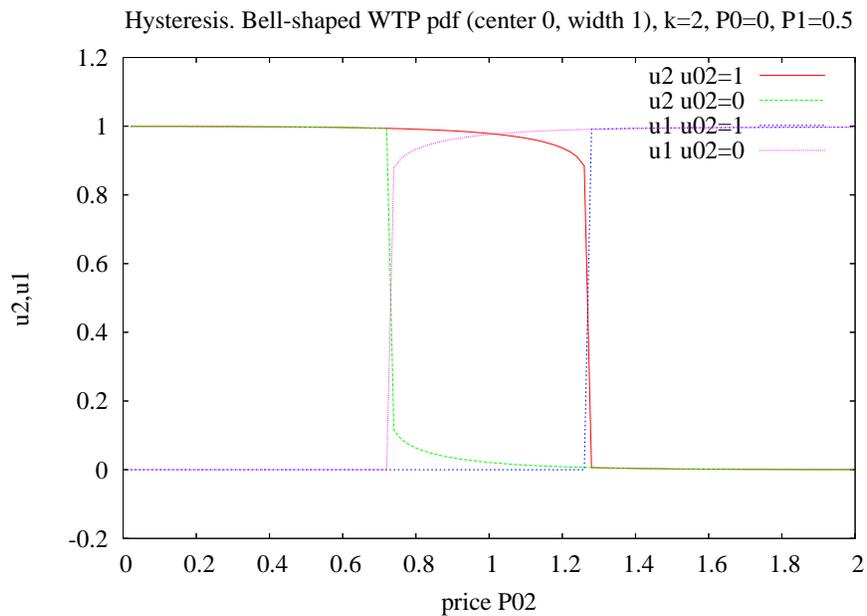}}
 \caption{A hysteresis cycle
obtained for $u_2$ when $P_{02}$ is varied. The red curve is $u_2$ for
$u_2(0)=1$ and the green curve when $u_2(0)=0$. The increasing returns
coefficient $k=2$ is larger than the WTP distribution width $w=4/\beta=1$. The
two curves coincide when the price $P_{02}$ is either small or large. They
differ in the intermediate price region.}
\end{figure}

The vertical transitions obtained in the parameter space for the attractor at
specific values of $P_{02}$ still correspond to finite time dynamics with a
characteristic time of $\lambda^{-1}$.

The middle part of figure 9 can be interpreted as follows. If we have a market
that is completely dominated by green cars, the maximum price $P_{02}$
has to rise almost up to 1.3 before the market is taken over by the intermediate or standard car.
On the opposite, if the green car does not have a substantial share of the
market, the price of the green car has to drop almost to 0.7 before the green
car can take over the market.

At the middle of hysteresis loop, $P_{02}$ is one width above the center of the
WTP distribution. But since $k=2$, the actual price when $u_2=1$ is one width
below the center of the distribution. Large values of $k$ imply a strong social
influence with respect to price differences: varying $u_2$ between 0 and 1
scans a large percentage of the WTP distribution (96 percent). At the
transition, $k=\frac{4}{\beta}=1$, this figure is already 63 percent.

\begin{figure}[htbp]
\centerline{\epsfxsize=120mm\epsfbox{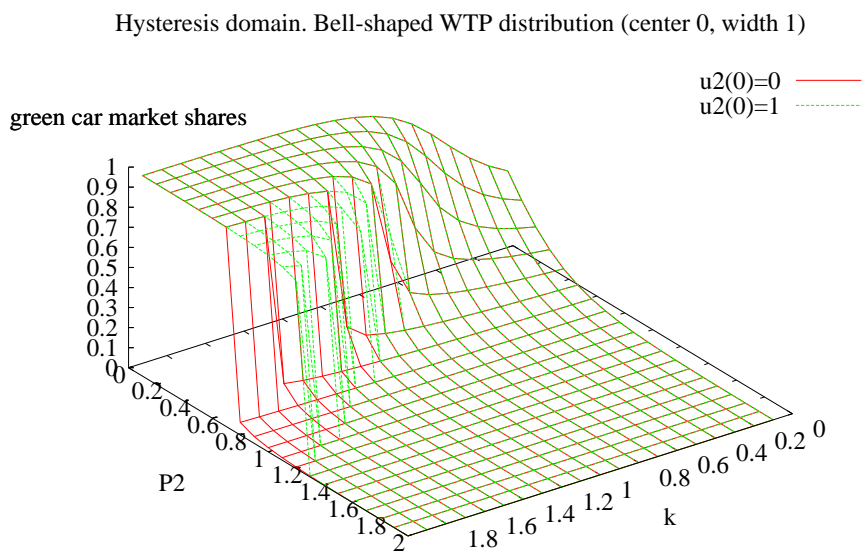}}
 \caption{Transition
between the hysteresis regime and the one attractor regime when
$k=\frac{4}{\beta}$. The green sheet is obtained when $u_2(0)=1$ and the red
sheet when $u_2(0)=0$. They differ only in the large $k$ and intermediate
$P_{02}$ region.}
\end{figure}

We only discussed until now the dynamics of $u_2$ as a function of $k$ and
$P_{02}$. Which of the two other products dominates or how they share the
market when $u_2=0$ depends upon the actual values of $P_{00}$ and $P_{01}$.

Why should we care about hysteresis? After all, our starting assumption was
that the intermediate and the green options are introduced after the standard
option implying $u_0=1, u_1=u_2=0$ as initial conditions. Since there are
several attractors in this regime, the issue of the adoption regime is
especially sensitive to the parameter set-up: if we now consider that
parameters can be under the influence of decision makers such as producers or
government agencies, or exogenous events (e.g., oil prices, technical
advances), the hysteresis regime can bring huge consequences for small
parameter changes. For instance,

\begin{itemize}
\item If price $P_{02}$ is lowered under the action of producers, advertising
in the media, or government subsidies, a transition from the $u_2 \simeq 0$
attractor to the $u_2 \simeq 1$ attractor can be induced.

\item Such an action does not have to be permanent: it might suffice to bring
$u_2$ above the separatrix, the central fixed point, to bring the system in the
basin of attraction of high $u_2$.

\item Competitors might also have equivalent strategies.

\item Sharp transition can be induced in this region through advertising by
decision makers, thus changing effective prices for some of the products or the
complete WTP distribution.
\end{itemize}

Multiple attractors are a challenge for scientists, but they are opportunities
for decision makers. Of course, parameter changes in the single attractor
regime also influence the outcome of the dynamics, but their influence is far
less dramatic. Moreover, these effects are reversed as the parameter changes
are undone, while the changes under multiple attractors might remain after
parameter changes are undone.

\section{Conclusions}
\label{conclusion}

In this paper, we have illustrated, how one can obtain interpretable but
already quite complex dynamics from a simple model on the competition between
more or less green technologies assuming that consumers have heterogeneous
preferences over these goods. In the first part of the paper we have shown
situations in which there was only one attractor for the dynamics. These
dynamics show that the prices of the greener alternatives needs to be far
enough from the boundary of the willingness to pay distribution to take over
the market as well as that the advantages of scale should be large enough to
overcome price differences over time. Depending on the precise parameters,
several different regimes are possible, differing in the number of equilibrium
technologies: 1, 2, or 3. In some situations, green technologies take over only
a smaller part of the market, while the standard technology remains dominant.
But there are also situations were first the intermediate technology conquers
some of the market, and thereafter the most green technology gains some market
share and takes most of the market.

Simulations were done with parameters being constant, by definition. One can
also infer the results of technological or attitude changes over market shares
dynamics. In situations in which there is basically only one attractor,
temporary policy measure will not have permanent results because the process
will reverse as soon as the policy measure ends.

In the second part of the results, we show that there are also situations in
which there are multiple attractors of the dynamics. They typically occur when
the width of the willingness to pay distribution is less than the increasing
returns coefficient. As long as the market share of the green products is
either very small (or very high)
the market is stable. However, if some agency is able to raise for some time
the fraction of consumers using green cars, or if it could boost for some period
the environmental consciousness of enough consumers, the system might jump 
to the situation in which most
people drive in green cars. Such temporary policy measures could then have a
 stable result even if the measure is only temporary. Of course
producers of standard cars could think of measures to get back to the first
situation, but these would at least be quite costly for them.

Let us stress that these dynamic properties are generic: they do not depend
upon a specific choice of the WTP distribution nor of the increasing return
price function. They apply to any S-shape WTP cumulative distribution and any
monotonic increasing return function.

Many extensions of the present model are possible, some more application
specific, others including coupling with pollution and opinion dynamics, the
role of government agencies etc. Some of these extensions might necessitate
heavier simulation tools such as multi-agent systems. But anyway, the simple
analysis that has been performed here already allows to figure out the
influence of the parameters on the observed dynamical regimes and the level of
behavioral complexity that can be expected for the heterogeneity of agents and
increasing return hypotheses.

Acknowledgments: We thank Jean-Pierre Nadal for illuminating discussions and
the participants of the CMAST07 workshop in Leiden (Feb-March 2007). G\'{e}rard
Weisbuch was also supported by E2C2 NEST 012410 EC grant. Vincent Buskens was
supported by the UU-High Potential Program `Dynamics of Cooperation, Networks,
and Institutions.'



\begin{thebibliography}{99}

\bibitem{art}
Brian Arthur. \emph{Increasing Returns and Path Dependence in the Economy}, Univ. of
Michigan Press, Ann Arbor, 1994.

\bibitem{jasss}
Fran\,{c}ois Bousquet, Robert Lifran, Mabel Tidball, Sophie Thoyer, and Martine
Antona. Agent-based modelling, game theory and natural resource management
issues. \emph{Journal of Artificial Societies and Social Simulation} 4(2),
2001, http://www.soc.surrey.ac.uk/JASSS/4/2/0.html.

\bibitem{gonaph}
Mirta B. Gordon, Jean-Pierre Nadal, Denis Phan, and Viktoriya Semeshenko.
Discrete Choices under Social Influence: Generic Properties,
http://halshs.archives-ouvertes.fr/halshs-00135405.

\bibitem{gonaphjv}
  Mirta B. Gordon, Jean-Pierre Nadal, Denis Phan, and Jean Vannimenus.
Seller's dilemma due to social interactions between customers. \emph{Physica A:
Statistical Mechanics and its Applications} 356(2-4), 15 October 2005, Pages
628-640.

\bibitem{kemp}
Ren\'{e} Kemp. The Diffusion of Biological Waste-water Treatment Plants in the
Dutch Food and Beverage Industry. \emph{Environmental and Resource Economics}
12, 113–136, 1998.

\bibitem{naphgo}
 Jean-Pierre Nadal, Denis Phan, Mirta B. Gordon, and Jean Vannimenus.
Multiple equilibria in a monopoly market with heterogeneous agents and
externalities \emph{Quantitative Finance} 5(6), December 2005, 557-568.

\bibitem{weisbuch2000}
G\'{e}rard Weisbuch. Environment and institutions: a complex dynamical systems
approach. \emph{Ecological Economics} 35(3), December 2000, Pages 381-391

\bibitem{kit}
Charles Kittel and Herbert Kroemer, \emph{Thermal Physics 2e}, 1980, W.H. Freeman
and Co, New York, 1980.

\bibitem{fol}
F\"ollmer H. (1974) "Random economies with many interacting agents",
{\it Journal of Mathematical Economics},  1, pp. 51-62.

\bibitem{fish}
 Weisbuch, G'erard, Alan P. Kirman and Dorothea K. Herreiner,
 Market organisation and trading relationships,
The Economic Journal, Vol. 110, n° 463, 411-436, 2000.

\end{thebibliography}
\end{document}